\documentclass[12pt]{article}
\usepackage{graphicx}
\usepackage{amssymb,amsmath}
\newcommand{\ZZ}{{\mathbf{Z}}}
\newcommand{\NN}{{\mathbf{N}}}
\usepackage{hfstyle}
\begin{document}

\title{Convergence to equilibrium in a class of interacting particle
systems evolving in discrete time}
\author{Henryk Fuk\'s
      \address{
         Department of Mathematics\\
         Brock University\\
         St. Catharines, Ontario  L2S 3A1, Canada\\
        \texttt{hfuks@brocku.ca}\\
         and\\
         The Fields Institute for Research\\
         in Mathematical Sciences\\
         Toronto, Ontario M5T 3J1, Canada
       }
      and Nino Boccara
      \address{
      Department of Physics, University of Illinois at Chicago,
      Chicago, IL 60607-7059, USA\\ \texttt{boccara@uic.edu}\\
      and\\ DRECAM/SPEC, CE Saclay,
91191 Gif-sur-Yvette Cedex, France}
   }

\Abstract{We conjecture that for a wide class of interacting
particle systems evolving in discrete time, namely conservative
cellular automata with piecewise linear flow diagram, relaxation
to the limit set follows the same power law at critical points. We
further describe the structure of the limit sets of such systems
as unions of shifts of finite type. Relaxation to the equilibrium
resembles ballistic annihilation, with ``defects'' propagating in
opposite direction annihilating upon collision.}

\maketitle
\section{Introduction}
Interacting particle systems evolving in discrete time have found
many applications in recent years, especially in the field of road
traffic modeling (see \cite{chow2000} and references therein).
While majority of traffic models based on interacting particle
systems are stochastic, some features of the real traffic flow may
be well described by purely deterministic models, as reported in
\cite{Nagel93,paper5,Nishi2000}.

An important characterization of a system of interacting particles
is its fundamental diagram, \textit{i.e.}, the graph of the
particle flux through a fixed point, in the equilibrium state, as
a function of particle density. In many cases, such diagrams
exhibit a discontinuity of the first derivative, and are very
often piecewise linear. In the simplest cases there is only one
such a discontinuity, and the fundamental diagram has the shape of
an inverted ``V''. Linear segments of the diagram can be
interpreted as distinct ``phases.''

The discontinuity of the first derivative, to be further called a
critical point, displays many features similar to critical
phenomena known in statistical physics. The phenomenon of critical
slowing down is especially apparent: when the density of particles
approaches the critical density, the convergence to the
equilibrium becomes slower. In what follows, we will present an
evidence which strongly suggests that at the critical point the
rate of convergence follows a power law with a universal exponent
equal to $-\frac{1}{2}$.

\section{Interacting particle systems and cellular automata}
Discrete version of a totally asymmetric exclusion process is a
simple yet often studied example of interacting particle
system.\cite{Krug88,Nagatani95,Nagel96,Sipper96,paper4} The
dynamics of this process can be described as follows: particles
reside on a one-dimensional lattice, with at most one particle per
site. At each time $t\in\NN$, each particle checks if the site to
the right of its current position is empty, and if it is, it jumps
to this site. The process is synchronous, meaning that all
particles jump at the same time.

Two alternative descriptions of this process are possible. We can
label each particle with an integer $n\in\ZZ$, such that the
closest particle to the right of particle $n$ is labeled $n+1$. If
$s(n,t)$ denotes the position of particle $n$ at time $t$, the
configuration of the particle system at time $t$ is described by
the increasing bisequence $\{s(n,t)\}_{n=-\infty}^{\infty}$. The
dynamics of the totally asymmetric exclusion process can be now
stated as
\begin{equation} \label{r184-00}
s(n,t+1)= s(n,t) + \min\{s(n+1,t) - s(n,t)-1,1\}.
\end{equation}

Another possible approach is to describe the process in the
language of cellular automata. If lattice sites are labeled with
consecutive integers $i\in\ZZ$, defining $x(i,t)=1$ if the site
$i$ is occupied by a particle, and $x(i,t)=0$ if it is empty, the
configuration of the particle system at time $t$ is, in this case,
described by the bisequence
$\{x(i,t)\}_{i=-\infty}^{\infty}$. One can easily show that the
dynamics of the aforementioned process is then given by
\begin{equation}
x(i,t+1)=f\Big(x(i-1,t),x(i,t),x(i+1,t)\Big),
\end{equation}
where $f$ is defined by
\begin{eqnarray} \label{r184-01}
& & f(0,0,0)=0,\,  f(0,0,1) = 0,\,  f(0,1,0) = 0,\, f(0,1,1)=1 \\
& & f(1,0,0)= 1,\, f(1,0,1) = 1,\, f(1,1,0)= 0,\, f(1,1,1) = 1.
\nonumber
\end{eqnarray}
The above definition can also be written in a more compact form as
\begin{equation} \label{r184-02}
x(i,t+1)=x(i,t) + \min\{x(i-1,t),1-x(i,t)\}
-\min\{x(i,t),1-x(i+1,t)\}.
\end{equation}
The evolution rule (\ref{r184-02}) is often called cellular
automaton rule $184$, using the numbering scheme introduced by
Wolfram \cite{Wolfram94}.

Consider now the general cellular automaton:
\begin{equation}
x(i,t+1)=f\big(x(i-r_l,t),x(i-r_l+1,t),\ldots,x(i+r_r,t)\big),
\label{g-rule}
\end{equation}
where the function  $f : \{0,1\}^{r_l+r_r+1}\mapsto\{0,1\}$ is the
evolution rule of the automaton, and the positive integers $r_l$
and $r_r$ are, respectively, the left and right radius of the
rule. $f$ will also be called a $k$-input rule where
$k=r_l+r_r+1$ is the number of arguments of $f$. A rule $f$ is
said to be conservative if for any periodic configuration of
period $L$ (\textit{i.e.}, a configuration such that
$x(i+L,t)=x(i)$ for every $i\in\ZZ$) we have
\begin{equation}
\sum_{i=1}^L x(i,t+1) = \sum_{i=1}^L x(i,t).
\end{equation}

Every conservative cellular automaton rule can be viewed as the
evolution rule of a system of interacting particles, just like
rule $184$ defined above, and a configuration of such a system can
be represented by an increasing bisequence
$\{s(n,t)\}_{n=-\infty}^{\infty}$, where $s(n,t)$ denotes the
position of particle $n$ at time $t$. A formal proof of this
statement and an algorithm for construction of an analog of
equation (\ref{r184-00}) for a given conservative rule $f$ can be
found in \cite{paper10}. In \cite{paper8} we have presented a
survey of 4-input and 5-input conservative cellular automata rules
and their properties.

In order to describe the motion of the particles, we will use a
visual representation of rule $f$ constructed as follows. List all
relevant neighborhood configurations of a given particle
represented by $1$. Then, for each neighborhood, indicate the
displacement of this particle  drawing an arrow joining the
initial and final positions of the particle. For rule 184, this
would be
\begin{equation}
\overset{\curvearrowright}{10},\quad
\overset{\circlearrowright}{1}1,
\end{equation}
where a circular arrow indicates that the particle does not move.
The above notation is equivalent to saying that all particles
which have empty sites immediately to their right jump one site to
the right, while other particles do not move.

We will normally list only neighborhoods resulting in particle
motion, assuming that by default, in all other cases, particles do
not move. Rule $184$, for example, we will simply represented by
$\overset{\curvearrowright}{10}$.

All particles do not necessarily move in the same direction. For
example, there could be a process such that a particle will move one site to
the right when the two nearest neighboring sites on its right are
empty while, if its right neighboring site is occupied and its
left neighboring site is empty, the particle moves one site to the
left. In all other cases, the particle remains immobile. This rule
can be visually represented by
\begin{equation}
\overset{\curvearrowright}{10}0,\quad
\overset{\curvearrowleft}{01}1.
\end{equation}

The main quantities of interest in this paper are the average
particle velocity and flux at time $t$. For a periodic
configuration of period $L$, the average velocity is defined as
\begin{equation}
v(t)=\frac{1}{N}\sum_{n=1}^{N}\big(s(n,t-1)-s(n,t)\big),
\end{equation}
where $N$ is the number of particles in a single period. The flux
is defined as $\phi(t)=\rho v(t)$, where $\rho=N/L$ is the density
of particles. In what follows, we assume that, at $t=0$, particles
are randomly distributed on the lattice. When $L\to\infty$, this
corresponds to the Bernoulli product measure on $\ZZ$, with
lattice sites occupied by a particle with probability $\rho$, and
empty with probability $1-\rho$. If $b=b_1b_2\ldots b_k$ is a
block of finite length $k$, where, for all $i\in\{1,2,\ldots,k\}$
$b_i\in\{0,1\}$, the probability of occurrence of block $b$ in the
configuration $\{x(i,t)\}_{i=-\infty}^{\infty}$  will be denoted
by $P_t(b)$.

\section{Exact solution for deterministic traffic rule}
A natural extension of the totally asymmetric exclusion process
with discrete time has been studied in connection with traffic
models \cite{Fukui96c}. Each site is either occupied by a
particle, or empty. The velocity of each particle is an integer
between 0 and $m$. As before, if $s(n,t)$ denotes the position of
the $n$th car at time $t$, the position of the next car ahead at
time $t$ is $s(n+1,t)$. With this notation, the system evolves
according to a synchronous rule given by
\begin{equation}
s(n,t+1)=s(n,t)+v(i,t), \label{FI1}
\end{equation}
where
\begin{equation}
 v(n,t)=\min\{s(n+1,t)-s(n,t)-1,m\}
 \label{FI2}
\end{equation}
is the velocity of the particle $n$ at time $t$. Since
$g=s(n+1,t)-s(n,t)-1$ is the gap (number of empty sites) between
particles $n$ and $n+1$ at time $t$, one could say that each time
step, each particle advances by $g$ sites to the right if $g \leq
m$, and by $m$ sites if $g>m$.  When $m=1$, this model is
equivalent to elementary cellular automaton rule $184$ discussed
in the previous section.

If we start with a random initial ($t=0$) configuration with
particle density $\rho$, it is possible to obtain an exact
expression for the flux at a later time $t>0$, as shown in
\cite{paper11}:
\begin{equation} \label{FIflux}
\phi(t) =1-\rho- \sum_{j=1}^{t+1} \frac{j}{t+1} {{(m+1)(t+1)}
\choose
 {t+1-j}} \rho^{t+1-j} (1-\rho)^{m(t+1)+j}.
\end{equation}
In the limit $t\to\infty$, we obtain the flux $\phi(\infty)$ at
equilibrium which is the piecewise linear function of $\rho$
defined by
\begin{equation} \label{FIfluxInfty}
\phi(\infty)= \left\{ \begin{array}{ll}
 m \rho  & \mbox{if $\rho<1/(m+1)$}, \\
 1-\rho    & \mbox{otherwise}.
\end{array}
\right.
\end{equation}
We can see that for $\rho<1/(m+1)$ the average velocity of
particles at equilibrium is $m$, \textit{i.e.}, all particles are
moving to the right with maximum speed $m$. The system is said to
be in the \emph{a free-moving phase}. When $\rho>1/(m+1)$, the
speed of some particles is less than the maximum speed $m$. The
system is in the so-called \emph{jammed phase}.

The transition from the free-moving phase to the jammed phase
occurs at $\rho=\rho_c=1/(m+1)$ called the \emph{critical
density}. At $\rho_c$, it is possible to obtain an asymptotic
approximation of (\ref{FIflux}) by replacing the sum by an
integral and using de Moivre-Laplace limit theorem, as done in
\cite{paper11}. At $\rho_c$ this procedure yields
\begin{equation}
\phi(\infty)-\phi(t) =\sqrt{\frac{m}{2 \pi (m+1)t}}
\left(e^{-\frac{(m+1)}{2 m t}} - e^{-\frac{(m+1) t}{2 m}} \right),
\end{equation}
or, in other words,  $\phi(\infty)-\phi(t) \sim t^{-1/2}$. Here,
by $f(t)\sim g(t)$ we mean that $\lim_{t\to\infty} f(t)/g(t)$
exists and is different from $0$. Power law convergence to the
equilibrium at the critical point for rule $184$ has been
established in \cite{Krug88} and \cite{Belitsky88}.

In the next section we will present some numerical results
suggesting that this behavior is universal for a wide class of
interacting particle systems evolving in discrete time.

\section{Numerical results for other CA rules}

Among conservative $5$-input cellular automata rules investigated
in \cite{paper8}, a majority exhibit piecewise linear fundamental
diagrams (\textit{i.e.}, graphs of $\phi(\infty)$ as a function of
$\rho$). For the purpose of this article, we have selected six
representative examples of such rules among the $428$ conservative
$5$-input rules. These six rules are given in Table 1, and their
fundamental diagrams are shown in Figure 1.
\begin{table}
\begin{center}
\begin{tabular}{|l|c|l|}   \hline
Rule nr & Motion representation \\  \hline 3213933712 &
$0\overset{\curvearrowright}{10}1$
$\overset{\curvearrowright}{11}0$
$1\overset{\curvearrowright}{10}$  \\ 3482385552  &
$0\overset{\curvearrowright}{10}1$
$0\overset{\curvearrowright}{11}0$
$1\overset{\curvearrowright}{10}$  \\ 3486567632  &
$0\overset{\curvearrowright}{10}1$
$1\overset{\curvearrowright}{10}$  \\ 3163339916  &
$1\overset{\curvearrowright}{10}$  \\ 3487613920  &
$0\overset{\curvearrowright}{10}0$
$1\overset{\curvearrowright}{10}$  \\ 3167653058  &
$00\overset{\curvearrowleft}{01}$
$11\overset{\curvearrowright}{10}$
\\ \hline
\end{tabular}
\end{center}
\caption{Examples of conservative rules with piecewise linear
fundamental diagrams}
\end{table}
Points at which the slope of the fundamental diagram changes are
referred to as critical points. In Figure 1, there are three rules
with a single critical point, and whose fundamental diagrams are
similar to rule 184. The remaining three rules have, respectively,
2,3, and 4 critical points.

In order to illustrate that at a critical point, as a function of
time $t$, the system converges to equilibrium as $t^{-1/2}$, we
define the decay time as
\begin{equation} \label{tau}
\tau = \sum_{t=0}^{\infty} |\phi(t)-\phi(\infty)|.
\end{equation}
If the decay is of power-law type, the above sum diverges.

For all rules in Table 1, we have performed computer simulations
to estimate $\tau$. Results are shown in Figure 2, where $\tau$ is
plotted as a function of density for each rule. The value of
$\tau$ has been estimated by measuring $\phi(t)$ for
$t=0,1,\ldots,1000$, and truncating the sum (\ref{tau}) at
$t=1000$.
\begin{figure}
\center{\includegraphics[scale=0.8]{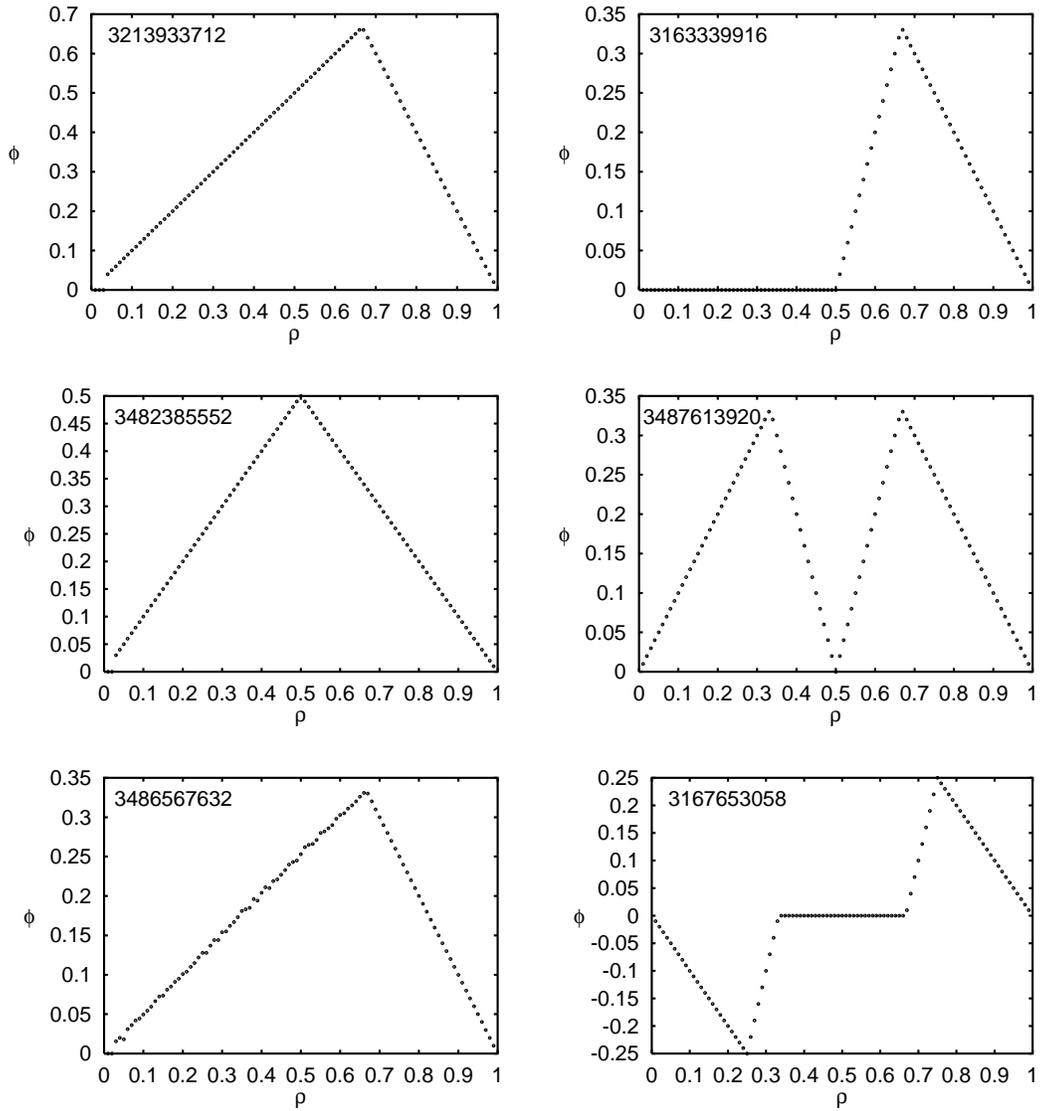}}
\caption{Fundamental diagrams for rules from Table 1.}
\end{figure}
\begin{figure}
\center{\includegraphics[scale=0.8]{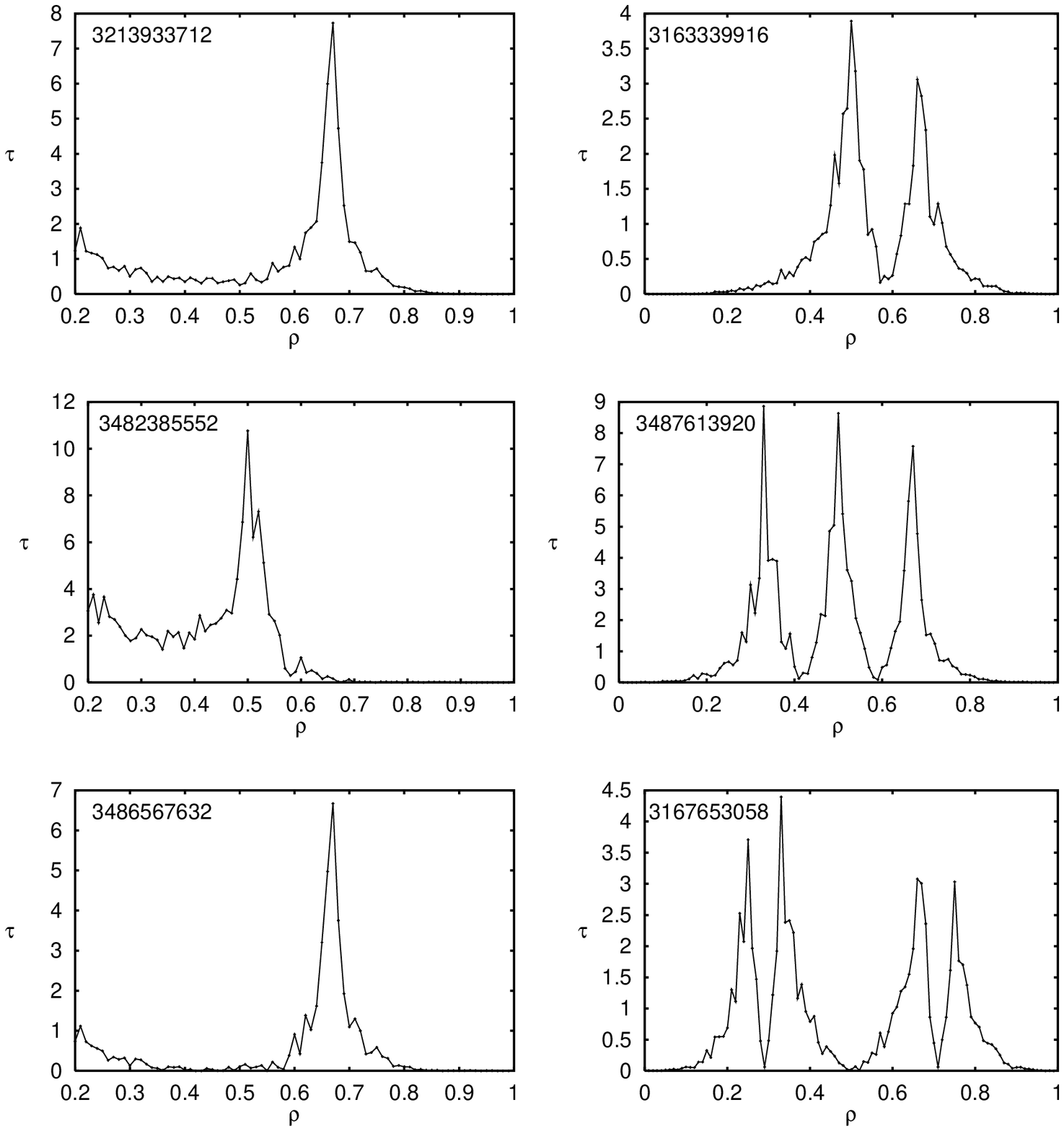}} \caption{Decay time
as a function of density for rules from Table 1.}
\end{figure}
Comparing Figures 1 and 2 we clearly see that $\tau$ diverges at
critical points. In order to verify that indeed
$|\phi(\infty)-\phi(t)| \sim t^{-1/2}$ at critical points, we have
plotted (for all critical points) the numerical values of
$|\phi(\infty)-\phi(t)|$  as a function of $t$ using logarithmic
coordinates. Figure 3 shows an example of such a graph, for the
critical point of the first rule in Table 1 (code number:
$3213933712$). The $(\rho, \phi)$ coordinates of this critical
point in the fundamental diagram are $(\frac{2}{3},\frac{2}{3})$.
\begin{figure}
\center{\includegraphics[scale=1.0]{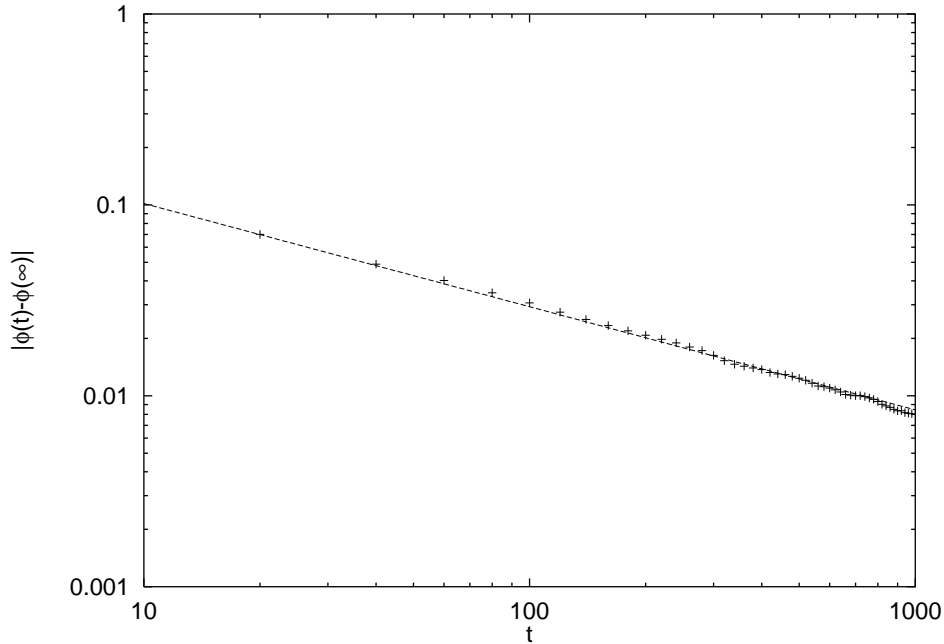}}
\caption{Logarithmic plot of $|\phi(\infty)-\phi(t)|$  as a
function of $t$ for rule $3213933712$. Data points ($+$) represent
computer simulations, while the dashed line represents the best
fit (slope $=-0.501\pm 0.01$)}
\end{figure}
Assuming that $|\phi(\infty)-\phi(t)| \sim t^{-\alpha}$, we have
determined the exponent $\alpha$ as the slope of the straight line
which best fit the logarithmic plot of $|\phi(\infty)-\phi(t)|$.
Table 2 shows our results for all the critical points of all the
six rules of Table 1. All these exponents are remarkably close to
$\frac{1}{2}$, suggesting that $t^{-1/2}$ might be a universal law
governing the rate of convergence to equilibrium at critical
points of conservative cellular automata with piecewise linear
fundamental diagrams.
\begin{table}
\begin{center}
\begin{tabular}{|l|l|l|} \hline
Rule nr & Critical point $(\rho, \phi)$ & $\alpha$ \\ \hline
\hline $3213933712$ & $(2/3,2/3)$      &  $0.501$ \\ \hline
$3482385552$ & $(1/2,1/2)$      &  $0.517$ \\ \hline $3486567632$
& $(2/3,1/3)$      &  $0.507$ \\ \hline $3163339916$ & $(1/2,0)$
&  $0.506$ \\ \cline{2-3}
             & $(2/3,1/3)$      &  $0.504$ \\ \hline
$3487613920$ & $(1/3,1/3)$      &  $0.502$ \\ \cline{2-3}
             & $(1/2,0)$        &  $0.539$ \\ \cline{2-3}
             & $(2/3,1/3)$      &  $0.504$ \\ \hline
$3167653058$ & $(1/4, -1/4)$    &  $0.486$ \\ \cline{2-3}
             & $(1/3,0)$        &  $0.514$ \\ \cline{2-3}
             & $(2/3,0)$        &  $0.514$ \\ \cline{2-3}
             & $(3/4,1/4)$      &  $0.486$ \\ \hline
\end{tabular}
\end{center}
\caption{Values of the exponent $\alpha$ at critical points of
rules from Table 1. Simulations performed using a lattice size
equal to $5\times 10^5$.}
\end{table}
\section{Shifts of finite type}
In order to understand why the decay to equilibrium, or, in other
words, approach to the limit set follows the same law for
different rules, we now describe limit sets of conservative rules
using the concept of a \textit{shift of finite type} (SFT). In
symbolic dynamics and coding theory \cite{Lind95}, SFT is defined as follows:
let $B$ be a finite set of finite blocks (consecutive sites),
\textit{e.g.} $B=\{00, 11, 101\}$. Consider a set of all
bi-infinite configurations in which blocks of the set $B$ do not
appear (``forbidden blocks''). Such a set, denoted by
$\mathrm{SFT}(B)$, is called a shift of finite type.

By analyzing frequencies of occurrences of finite blocks in
conservative CA, one can observe that limit sets of many
conservative CA are unions of two or more SFT's. Each of these
SFT's correspond to a distinct "phase", or a straight line segment
in the fundamental diagram. Critical points correspond to a set of
configurations common to two SFT's. Forbidden block sets of each
phase completely determine the fundamental diagram.

As an example, consider the deterministic traffic rule defined by
the relations (\ref{FI1}) and (\ref{FI2}), assuming $m=2$. We
mentioned that in the free-moving phase, all particles move with
maximum speed $m=2$, which implies that 1's are always separated
by two or more 0's. Hence, the blocks $11$ and $101$ cannot be
found in a configuration belonging to the limit set of the
free-moving phase. In the jammed phase, on the other hand, blocks
of zeros of length 3 or more cannot occur. The limit sets of these
two phases are therefore, respectively, $\mathrm{SFT}({11, 101})$
and $\mathrm{SFT}({000})$. Note that these two sets are not
completely disjoint, but have three common elements, namely,
$...100100100100...$, $...001001001001...$, and
$...010010010010...$

Consider now the phase $P1=\mathrm{SFT}({11, 101})$. It is clear
that, for any configuration in $P1$, $\rho\leq1/3$, since two ones
must be always separated by two or more zeros. We know that the
flow for the above rule is defined as $\phi(t)=2P(100)+P(101)$. We
also know that $P(11)=P(101)=0$. The consistency condition gives
$P(100)+P(101)+P(110)+P(111)=\rho$. Hence, $P(100)=\rho$, and, in
this phase, $\phi(\infty)=2\rho$.

In phase $P2=\mathrm{SFT}({000})$, $P(000)=0$, and one can show
that $\phi(\infty)=1-\rho-P(000)$. The proof can be found in
\cite{paper11} (it involves manipulations of the consistency
conditions only). Hence, $\phi=1-\rho$. In $P2$, the largest
number of zeros separating two ones is $2$, hence the minimum
value of the density $\rho$ is $\frac{1}{3}$, so
$\phi(\infty)=1-\rho$ is valid for all $\rho\ge\frac{1}{3}$.

Note that we derived the fundamental diagram given by equation
(\ref{FIfluxInfty}) from the structure of the limit set only. This
can be done for all conservative CA with piecewise linear
fundamental diagrams. As mentioned earlier, different straight
line segments of the fundamental diagram correspond to different
SFT components of the limit set, while configurations common to
two SFT's define critical points.

At the critical point, the limit set consists of a finite number
of configurations, yet there is an infinite number of initial
configurations with a given density. Time evolution of the system
governed by a conservative rule can, therefore, be viewed as a
transition form an infinite configuration space to a finite
configuration space. While details of this transition are
different for different rules, it appears that the rate of
convergence follows always the same law, as argued in the previous
section.

In order to test this interpretation of the dynamics of
conservative CA, we have determined the spatial measure entropy of
configurations at time $t$. The spatial measure entropy of a limit
set configuration is defined as
\begin{equation}
s(k)=-\frac{1}{k}\sum_{b \in \{0,1\}^k} P_{\infty}(b) \log
P_{\infty}(b),
\end{equation}
where the sum runs over all blocks of length $k$, and
$P_{\infty}(b)$ denotes the probability of occurrence of block $b$
in the limit set. If, at a given $\rho$, the number of
configurations in the limit set is finite, $s(k)$ should go to
zero with $k\to\infty$. Figure 4 shows $s(10)$ plotted as a
function of the density $\rho$ for all the six rules of Table 1.
Comparing Figures 1 and 4, one can easily notice that, at critical
points of fundamental diagrams, $s(10)$ takes minimum values,
confirming our observations regarding the structure of limit sets.
\begin{figure}
\center{\includegraphics[scale=0.8]{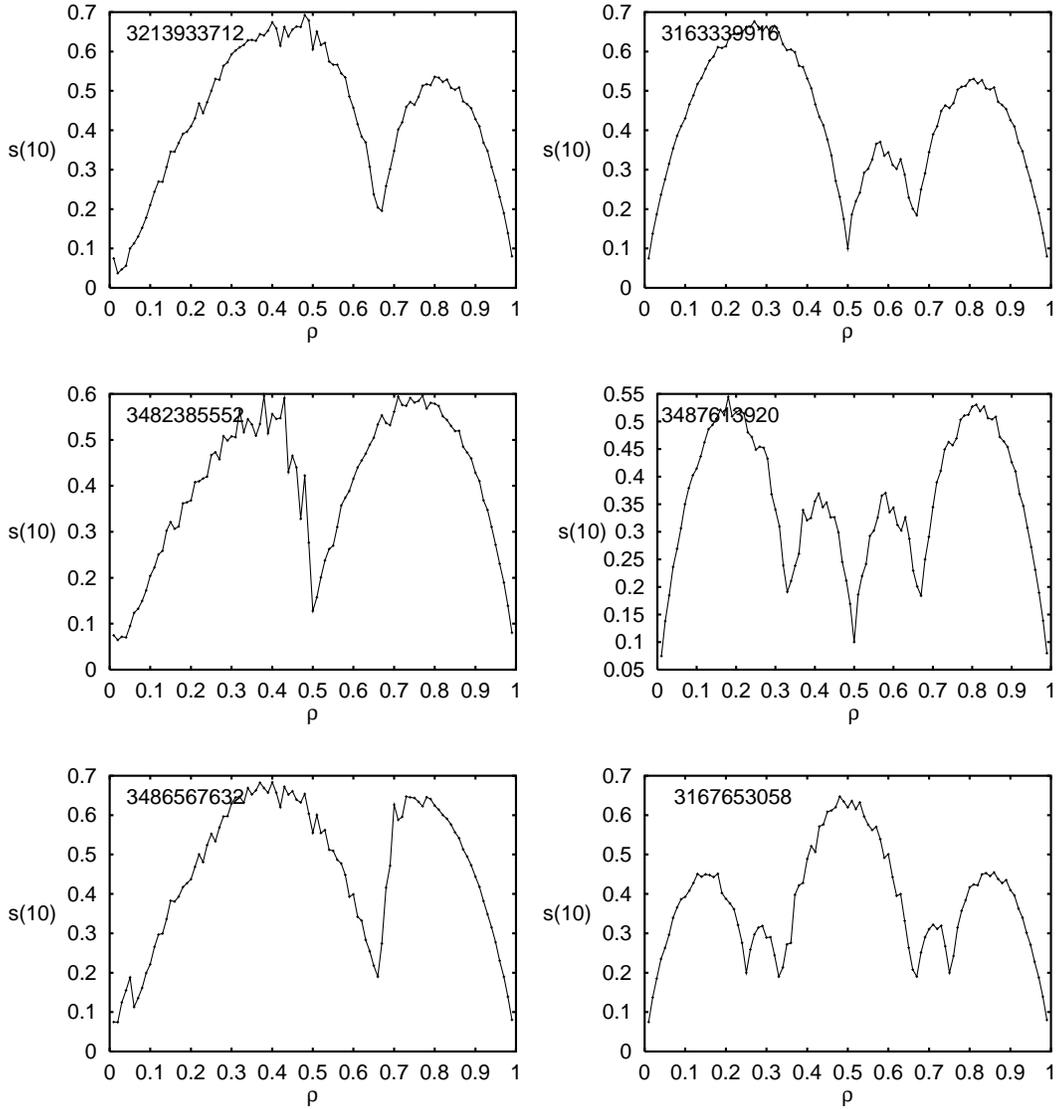}} \caption{Spatial
measure entropy for block length $k=10$ as a function of density
for rules of Table 1.}
\end{figure}
\section{Annihilation of defects}
The dynamics of rule 184 can be described in terms of soliton-like
localized structures propagating in periodic background
\cite{BNR91} More formally, it can be shown that rule 184 is
equivalent to ballistic annihilation process \cite{elskens85}.
This equivalence has been used in \cite{paper4}, and investigated
in detail in \cite{Belitsky88}. The ballistic annihilation process
involves two types of particles moving, on a line, at constant
speed in opposite directions. When two particles collide, they
annihilate. Using the central limit theorem, Elskens and Frisch
\cite{elskens85} demonstrated that if the initial distribution of
particles is balanced, then, the fraction $S(t)$ of surviving
particles  at time $t$ behaves  as $t^{-1/2}$.

For rule 184, the ``background'' on which the defects are moving
are periodic configurations of alternating zeros and ones, $\ldots
01010101 \ldots$. If we call $A$-type defect consecutive 1's, and
$B$-type defect two consecutive 0's, it is easy to show
\cite{paper4} that, for each iteration of rule 184, defects type
$A$ move to the left, while defects of type $B$ move to the right.
When they collide, the reaction $A + B \rightarrow background$,
\textit{i.e.}, annihilation, takes place.

A very similar construction can be made for the process defined by
equation (\ref{FI1}) and (\ref{FI2}). Here, the background
consists of ones separated by $m$ zeros. We can also define two
types of defects, but with an additional subscript identifying
their length. If two consecutive 1's are separated by a cluster of
zeros of length $m+k$ ($k>0$), this is the defect of type $B$,
denoted $B_k$. Similarly, if two 1's are separated by a cluster of
zeros of length $m-l$ ($m> l>0$), such a cluster constitutes an
$A$-type defect, denoted $A_l$. $A$-type defects move to the left
with speed $1$ while $B$-type defects move to the right with speed
$m$. A collision between defects $A_l$ and $B_k$ results in a
defect $A_{l-k}$ if $l>k$, and $B_{k-l}$ if $k>l$. If $k=l$,
the two defects annihilate. Figure 5 shows an example of such
collisions.
\begin{figure}
\center{\includegraphics[scale=1.0 ]{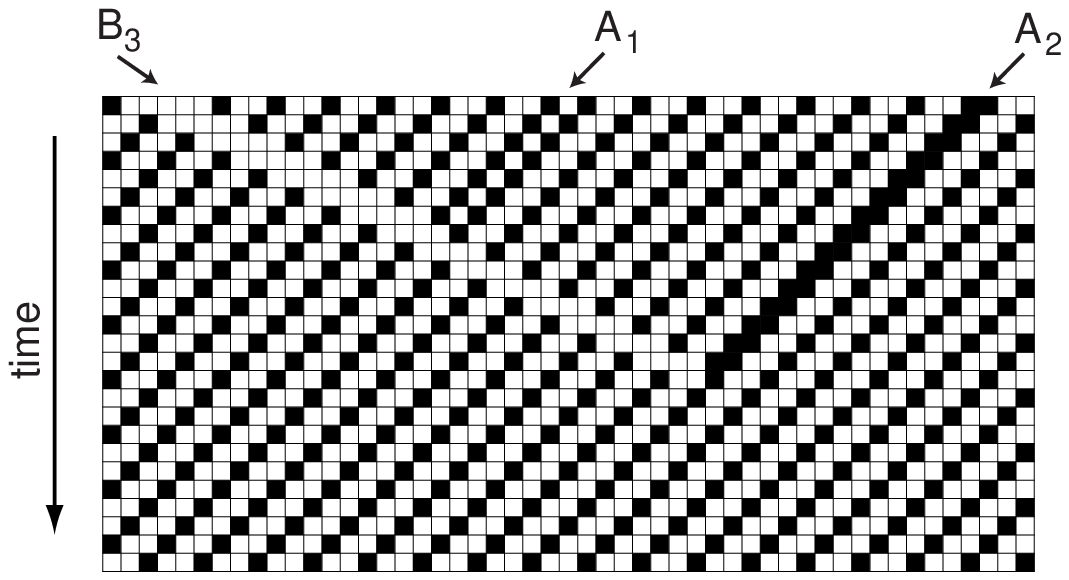}} \caption{Collision
of defects in deterministic traffic rule defined by relations
(\ref{FI1}) and (\ref{FI2}) with $m=2$. Black squares represent
lattice sites occupied by a particle. First, $A_1$ collides with
$B_3$, resulting in $B_2$. Several iterations later, $B_2$
collides with $A_2$, and both defects disappear.}
\end{figure}

The dynamics of other conservative rules with piecewise linear
fundamental diagrams can also be interpreted as interaction of
defects propagating in a periodic background. However, the number
of different types of ``defects'' and ``backgrounds'' is typically
larger than in the simple cases described above. This can be seen
in Figure 6, which shows spatiotemporal diagrams for rule
$31633399164$ ($1\overset{\curvearrowright}{10}$) at
$\rho=\frac{1}{2}$ and $\rho=\frac{2}{3}$. ``Defects'' propagating
in opposite directions and annihilating upon collision can be
clearly identified in both diagrams. At $\rho=\frac{1}{2}$ all
defects eventually disappear resulting in  a periodic
configuration $\ldots 01010101 \ldots$. Similarly, at
$\rho=\frac{2}{3}$ annihilation of all defects results in a
periodic configuration $\ldots110110110\ldots$. Since these
processes can be considered a generalization of a simple ballistic
annihilation, one can conjecture that their rate of convergence to
equilibrium should be the same as for the ballistic annihilation,
in agreement with the results of computer simulations presented in a previous section.
\begin{figure}
\center{a)\includegraphics[scale=1.0]{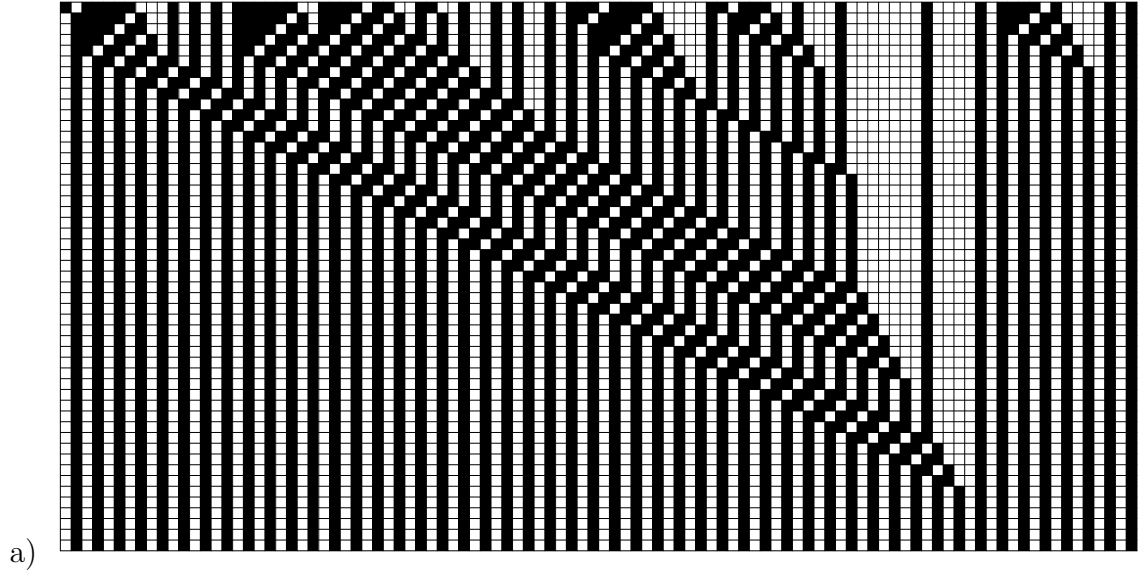}}\\
\center{b)\includegraphics[scale=1.0]{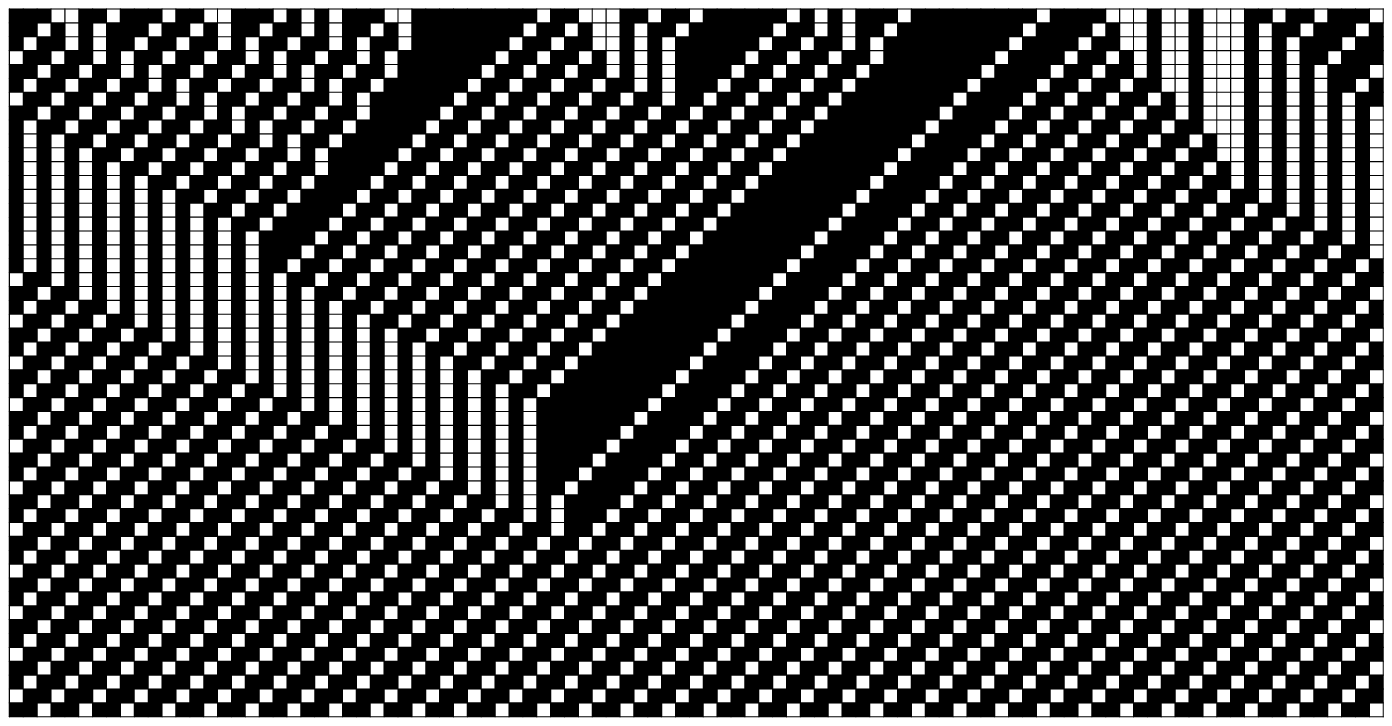}}
\caption{Spatiotemporal diagrams for rule $31633399164$
with density of particles 
a)$\rho=\frac{1}{2}$ and b) $\rho=\frac{2}{3}$.}
\end{figure}

\section{Conclusion}
We have presented numerical evidence suggesting that, for
conservative cellular automata with piecewise linear flow
diagrams, relaxation to equilibrium  at a critical point follows a
universal power law with exponent $-\frac{1}{2}$. The universality
of this critical behavior is related to the structure of limit
sets of such rules, which can be described as unions of shifts of
finite type. At critical points the spatial measure entropy of the
limit set goes to zero, which means that the limit set consists of
a finite number of configurations. Moreover, the dynamics of such
rules can be viewed as interaction of ``defects'' propagating in
opposite directions and annihilating upon collision, just like in
a simple ballistic annihilation process, for which the power law
$t^{-1/2}$ behavior has been established analytically.

\end{document}